# Hypermedia Learning Objects System - On the Way to a Semantic Educational Web

*M. Engelhardt, A. Kárpáti, T. Rack, I. Schmidt, T.C. Schmidt*

Fachhochschule für Technik und Wirtschaft Berlin

**Abstract:**

*While eLearning systems become more and more popular in daily education, available applications lack opportunities to structure, annotate and manage their contents in a high-level fashion. General efforts to improve these deficits are taken by initiatives to define rich meta data sets and a semantic Web layer.*

*In the present paper we introduce Hylos, an online learning system. Hylos is based on a cellular eLearning Object (ELO) information model encapsulating meta data conforming to the LOM standard. Content management is provisioned on this semantic meta data level and allows for variable, dynamically adaptable access structures. Context aware multifunctional links permit a systematic navigation depending on the learners and didactic needs, thereby exploring the capabilities of the semantic web.*

*Hylos is built upon the more general Multimedia Information Repository (MIR) and the MIR adaptive context linking environment (MIRaCLE), its linking extension. MIR is an open system supporting the standards XML, Corba and JNDI. Hylos benefits from manageable information structures, sophisticated access logic and high-level authoring tools like the ELO editor responsible for the semi-manual creation of meta data and WYSIWYG like content editing.*

## 1 Introduction

Building educational content significantly depends on the target media: In writing a book we create an unchangeable, monolithic block of strict linear order. Setting up a collection of HTML documents results in a mesh of easily changeable content elements, which show a strict page orientation. The relational mesh itself, when fixed with the rather rigid HTML linking scheme, withstands any seamless modification. Most of the learning platform environments on the market today follow the latter framework, thereby inheriting its severe shortcomings. Authors as well as consumers thus are forced to cope with material inherently shaped by the publication method.

The early days of hypertext already brought up alternative ideas of online content forms: Self-consistent information objects, possibly consisting of several connected components, should serve as input for runtime environments more elaborate than today's browsers [14]. Information fragments were foreseen to be loosely coupled by relational hyper reference components, offering impressive flexibility and the perspective of augmented interpretation. Retrospectively examined such pioneer work as the Dexter Model [6] required complementation by three major steps: The notion of context had to be introduced at the conceptional level, at first. Presentation independent encoding techniques were needed for content and hyperlinks, which appeared with the XML family [2]. Finally, definitions of structural standards for information encoding and content organisation had to be invented.

In education the latter task has been addressed lately with the emerging standards Learning Object Meta data (LOM) [9] and Sharable Content Object Reference Model (SCORM) [10]. LOM introduces an annotation hierarchy of meta data of intermediate complexity, but also the notion of learning objects as a collection of content components together with its meta data. LOM's ELearning Objects (ELOs) revitalise the idea of rich, coherent information entities, subject to an appropriate processing for presentation. LOM itself forms one ingredient of the fairly comprehensive SCORM concept, which suffers though from limited practicability due to its normative implementation instructions of inferior technical kind.

Building educational content using ELOs instead of HTML-pages opens up a variety of fascinating new opportunities for authors, teachers and learners. At the same time this concept imposes some specific restrictions and open questions, which are currently under lively discussion [8]. On the one hand variable content access paths and online views may be generated from the same collection of re-usable ELOs, individually adapted to specific contexts of teachers and learners. On the other hand authors need to provide the required meta data and have to cope with the complexity of breaking content into self consistent units. In such granular content concepts both authors and readers may have to struggle with identifying a coherent train of thought.

Coherence in modular hypermedia information systems is mainly gained from hyperlinks. Well prepared navigational intelligence should be regarded as a key issue, as it glues together parted knowledge in semantic terms. For any linking scheme it is therefore desirable to follow strong demands on consistency and transparency, a rhetoric pattern for example as stated in the fundamental work of Landow [1]. A rigorus attention of semantic rules in link production and selection cannot be expected without appropriate support of the authoring and management environment. The Semantic Web initiative has raised new ideas and techniques to approach this goal in a flexibly automated, high level fashion.

In the present paper we introduce our prototypic solution of an ELO based open hypermedia system, the Hypermedia Learning Object System HYLOS [3], donating special focus to an efficient authoring of ELOs. Furthermore we discuss semantic interpretations of interactivity in this context. Our approach starts from the primary, most often violated principle of educational content applications, the strict separation of structure, logic, content and design, as it can be achieved by applying XML-technologies in a rigorous fashion. Here it should be noted that hyperlinks, from our view, belong to structural information and therefore must not be stored within content. We will discuss a semantic representation of hyperlinks and a prototypic model for a semantic link processing along the line of this article.

This paper is organised as follows. Section 2 presents the HYLOS system and its basic concepts. In section 3 we introduce a translational scheme of metadata decorated content into semantic statements, including a representation of hyperlinks and their contextual processing. Finally, section 4 gives a conclusion and an outlook on the ongoing work.

## 2  The Hypermedia Learning Object System HYLOS

### 2.1  Managing ELO Content

In this section we want to introduce HYLOS, our model and prototypic implementation of an educational content management solely built on eLearning Objects. Operating on a base of ELOs HYLOS pre-processes content for variable views: Each component can be displayed as comprehensive slide or detailed descriptive information. According to LOM attributes the learning complexity and the semantic density of all presentations may be adjusted. Different access structures are provided according to the didactic model in use. Figure 1 shows the hierarchical instruction path representing a behaviouristic approach, a set of individualised tools in constructivist fashion for searching, navigation and interactively aided overviews on completeness and learning success is under way. HTML and PDF currently are supported as presentation formats, where other types of preparation may be added easily.

Hyper referential relations within this applications are adaptable by authors as well as by users on a semantic layer. In HYLOS an author can define contexts of hyperlinks, representing the rhetoric of his choices. Learners may opt for the link context in use for their navigation. These capabilities are the outcome of the MIR adaptive Context Linking Environment described below, which is part of hypermedia learning system.

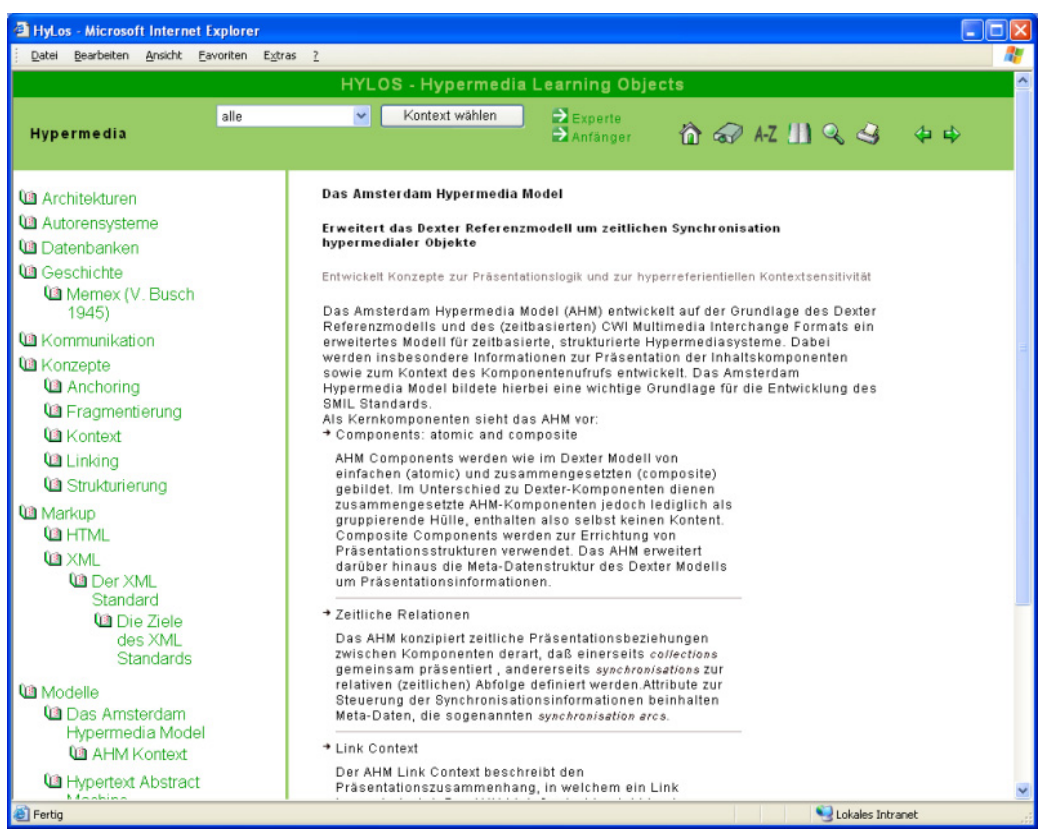

Figure 1: Instructional Hierarchy – one Access Path in HYLOS

As proposed by the LOM standard ELOs in our system are simultaneously formed from content entities and meta descriptors. Content is built from XML paragraph objects, as is the standard cellular content concept within the underlying MIR system [13]. Meta data are selected as a conformal subset of the LOM standard. Additional information such as taxonomies according to external categorisation schemes, glossaries, bibliographies or organisational data have been modeled within the system and become accessible to ELOs by reference identifiers. The complete information model is displayed in figure 2.

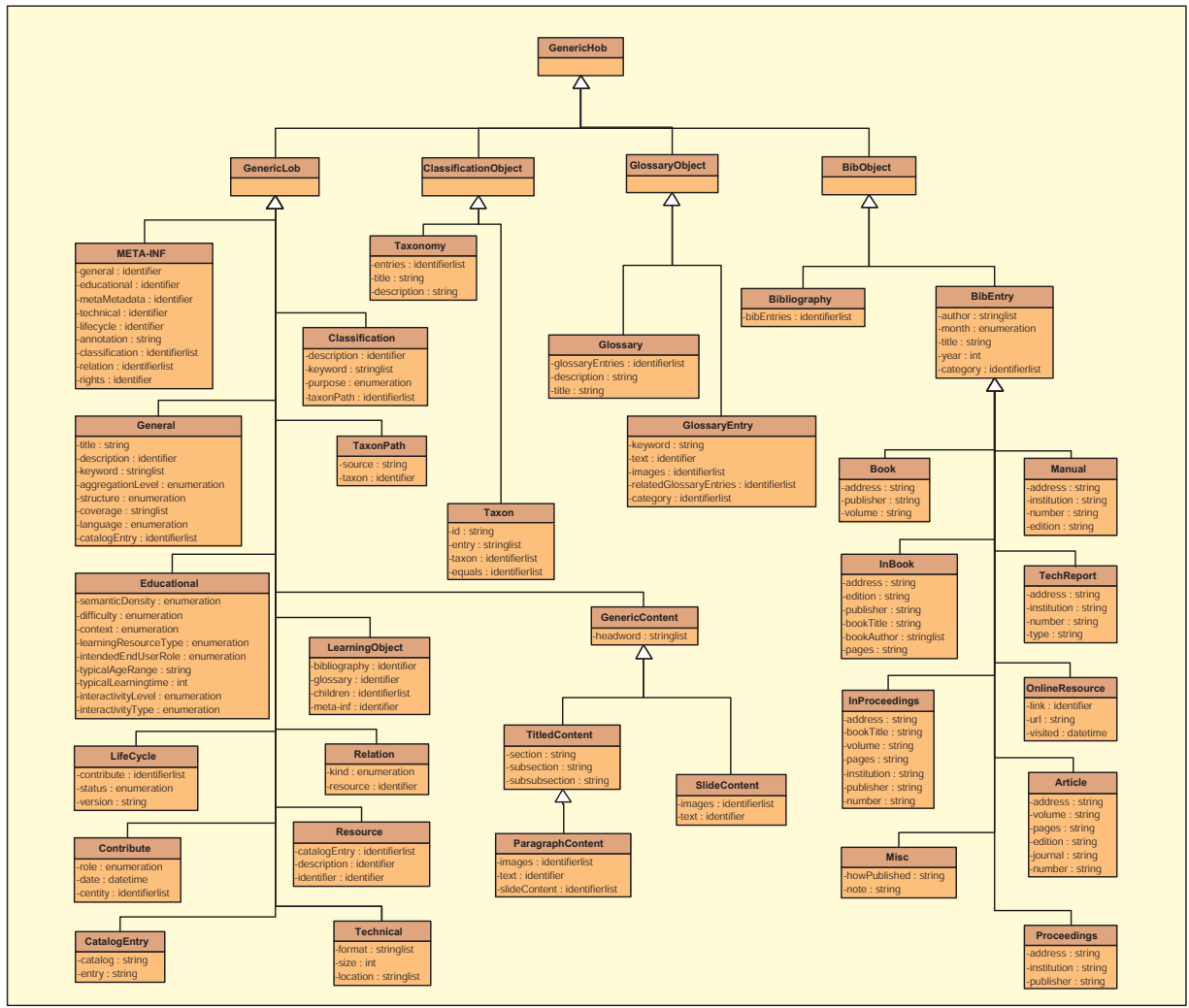

Figure 2: Class Diagram of the ELO Information Model

Our practical implementations of the HYLOS system rank around XML formats and processing technologies. They rely on the more general storage and runtime platform Multimedia Information Repository (MIR) [3]. Grounded on a powerful media object model MIR was designed as a universal fundament for ease in modelling and implementing complex multimedia applications. All data residing in the adaptable MIR data store are published in XML format, such that individual views and user interface behaviour can be reached by lightweight style sheet programming.

Built on a three-tiered architecture MIR provides general support of media data handling, authentication, user and connection handling. Its core is formed by a media object database, implementing a duality of object oriented information model and relational structure. The system offers a free layer for application specific modelling of information and structures, the latter being twofold as passive structures and 'active' references, where traversal is accompanied by underlying code execution. A generic web authoring allows for immediate editing of the modelled information and structures. MIR is built as an open hypermedia system and currently supports the standards XML, Corba and JNDI. For further reading we refer the reader to [4] and [5].

### 2.2  Efficient ELO Authoring

Authoring Learning Objects is not a simple task: Content has to be comprehensively shaped for covering a single, self-consistent subject. Meta data, in a certain amount, are inevitably needed. It is a necessary but ambitious challenge to provide an authoring tool for seamless production of ELOs.

The HYLOS ELO editor (s. fig. 3) allows for a coherent authoring of complete learning objects, i.e. content, meta data and referential relations can be developed within *one* application. The tool attains three main views: The content navigator, the content editor and the meta data builder.

The **content navigator** offers the traversal and modification of ELO structures, operating on the relational context paths described in [13]. Note that, as the applicative ELO structure need not be hierarchical, the generated view of an object tree is in the case of object re-use a non-normalised representation of the content.

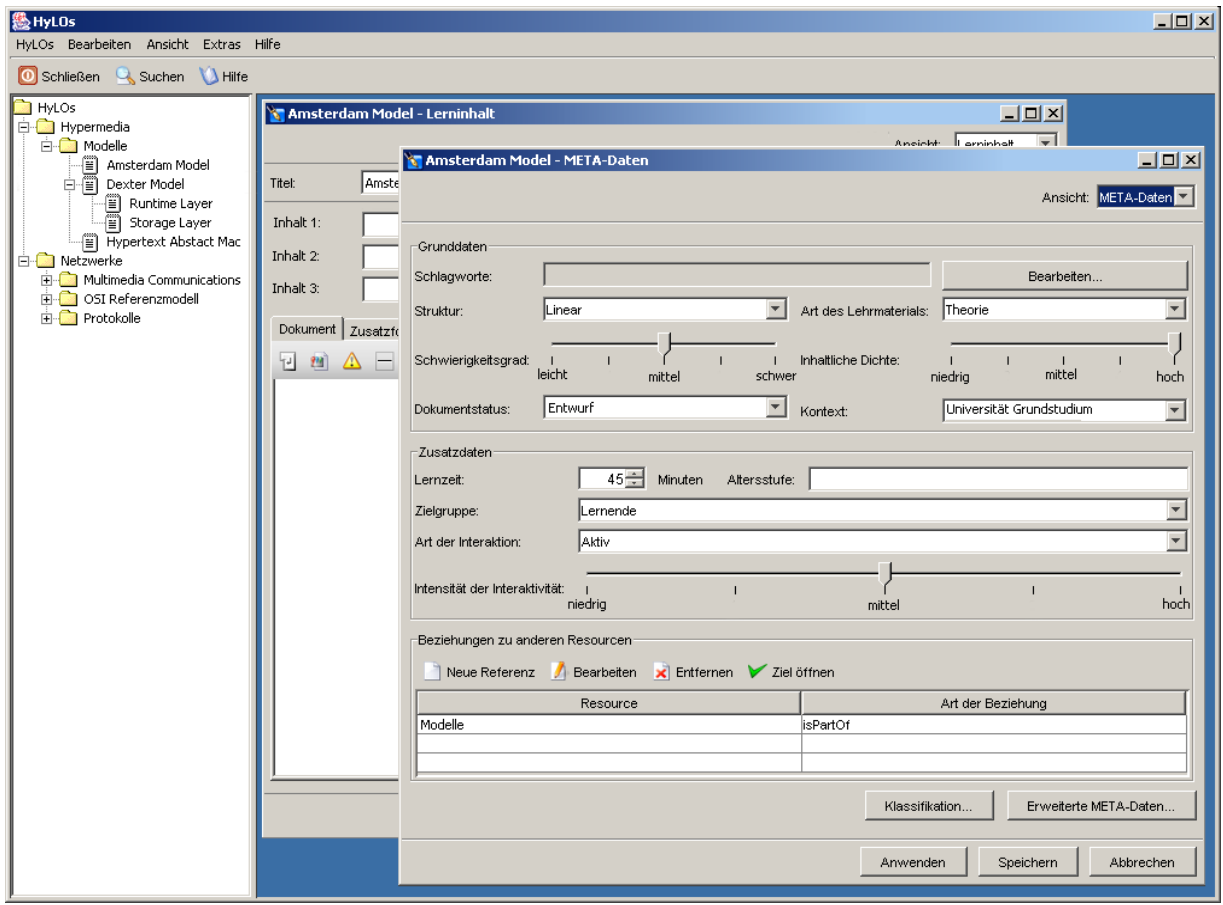

Figure 3: The HYLOS ELO Editor

The **content editor** is dedicated to the production of the entire content, i.e. descriptive paragraphs and slides. The main information structure to be filled is the ParagraphContent (s. fig. 2), consisting of an XML-formatted free text paragraph (including images or other media) and descriptive elements s.a. title, headwords and sectional titles. The latter strings are recycled to automatically generate a 'standard' slide for every ELO for the 'quick and simple' slide production. For voluntary use HYLOS offers an unrestricted slide presentation layer within the SlideContent information object, which can be correspondingly authored with the XML paragraph editor.

The **meta data builder** takes care of generating the ELO meta data set with minimised authoring effort. Relevant manual specifications are arranged on *one* sheet, where obligatory data are reduced to *seven* fields at the upper part (s. fig. 3). The acquisition of meta data is essentially done in three ways:

- o **Automatic Generation** for most of the LOM attributes: All technical data (author, formats, sizes, dates, locations, aggregationLevel …) are directly provided by the MIR system. The content title is used as the LOM title, the sectional titles as coverage fields and as a description the (reformatted) first content paragraph. An additional set of faintly fluctuating data, e.g. language or intendedEndUserRole, are taken from user specific presets.
- o **Obligatory manual provision** for seven LOM attributes: Keywords, semanticDensity, difficulty, context, learningResourceType, structure and documentStatus require editing, if presets taken from previous editing do not apply.
- o **Facultative manual provision** for the remaining LOM attributes may be added either on the front sheet or by accessing the complete meta data tree.

Additional information structures s.a. glossary entries, taxonomic classifications, bibliography entries or persons can be accessed within the ELO editor through separate window sheets. Thus an author of ELOs is enabled to create or manipulate complex objects without distracting the focus from its destination in content.

# 3 MIRaCLE – an Adaptive Linking Environment

## 3.1 Identifying the Semantic of Hyperlinks

Assuming LOM metadata in presence at content items a canonical semantic description is easily derived: Using RDF [7] representation the content object attains the role of the subject, the name of the meta descriptor forms the predicate and the value of it denotes the object. For example a LOM general description "about hamster diseases" would turn into the statement "this learning object is a description about hamster diseases" in RDF.

To approach a semantic analysis of hyper referential links let us recall that a hyper reference is constructed of two entities, anchors and links. Links concatenate anchors, which identify sub portions of content. In a fairly general fashion anchors can be expressed within XLink statements by XPointer/XPath-like expressions [2], the exact formalism depending on the media type of the document. Links as well as anchors may be stored separate from document resources, e.g. in a link base.

Even though it appears rather straight forward that a semantic description of an anchor should inherit the expository statements of the underlying content, sole information inheritance remains insufficient, since a document in general may carry several, sub specific anchors. It is therefore important to provide additional specifications as can be done by the title and label tags inherent with XLink locator expressions. Note that the denoted data chunks in anchors need not be of textual type. Anchors in this sense must be viewed as additional specialisations, i.e. "this resource in the context of hamster diseases carries the title of hamster having hay fever". The extraction of a semantic description of anchored resources given as a collection of inherited and dedicated statements is visualised in figure 4.

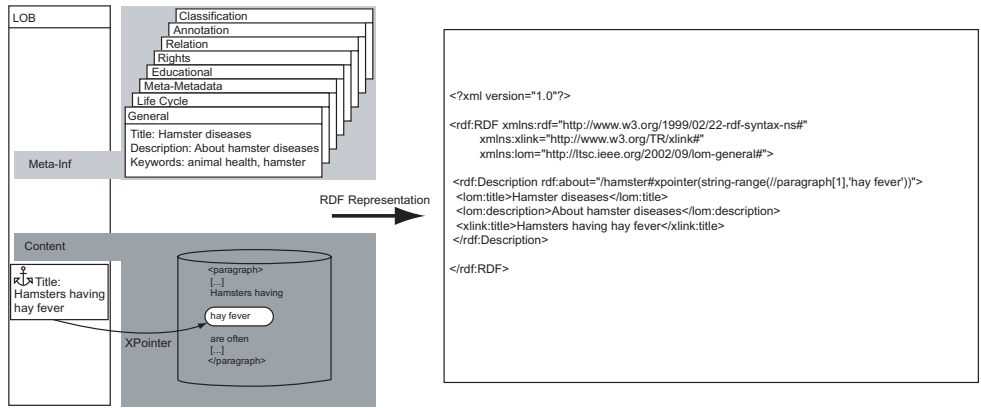

Figure 4: Gaining Additional Anchor Descriptors

Links denote relations between two or more anchors. They are directional components, uni- or bi-directional. Following the XLink arc encoding a link expression itself may carry directionless attributes, multiple titles, as well as directed descriptors, e.g. the arc attributes from, to and arcrole. A semantic of hyperlinks then naturally should build up on matching attributes, i.e. using arcroles whenever arc and direction apply, and on the linked resources. At first, this gives rise to a collection of simple statements: "This link carries the title 'For freshman'", "This link starts from the resource 'hamster having hay fever'", etc.

In semantic terms statements represent linked resources. A link encodes a relation between them, which is directionally attributed by means of the arcrole. Thus at second an XLink expression gives rise to a more complex, reifying statement. A link expresses via its arcrole attribute a predicate describing the referred resources. However, in transforming this notion into a simple statement, the link resource itself remains unseen.

To cure this deficit a higher order statement, a statement about statements, needs to be used. Following this approach the link entity forms the subject for a statement about this relation description statement. As is visualised in figure 5 such expression reads, "Link1 denotes that resource 'Hay fever handbook' presents BackgroundInfo to resource 'Hamsters having hay fever'".

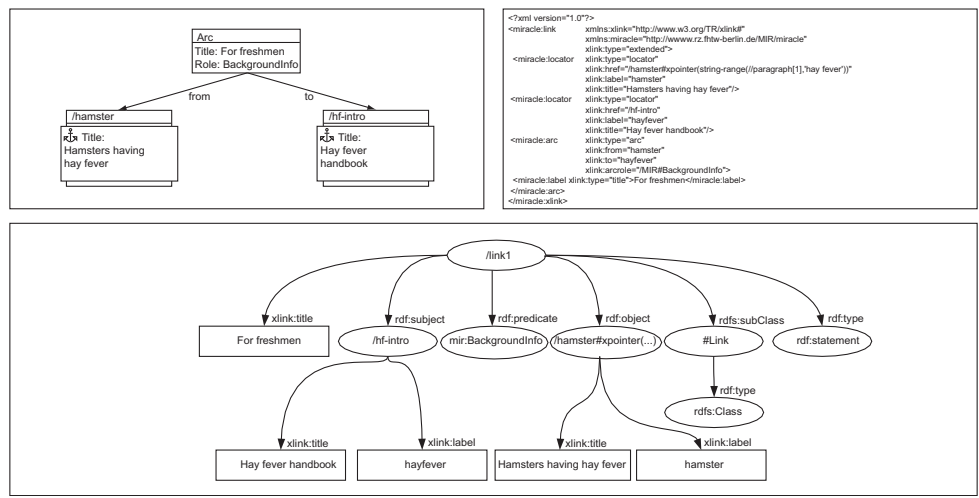

Figure 5: From XLink to an RDF Hyperlink Description

Expressing the core semantics of hyperlinks as higher order statements opens the opportunity to preserve relation to contextual information s. a. link titles etc. Viewing the approach in a rigorous semantic fashion this is indeed correct, as a link may form a resource external to content, its denoted relation being not true by itself, but an expression of contextual and personal view of the (link) author, who may be distinguished from content authors.. For further reading we refer to [13].

## 3.2 Semantic Link Context

Connecting distributed knowledge resources is more than simply adding a link to a document. Having derived a semantic notion of annotated content, anchors and hyperlinks in section 3.1 we are now able to define a high-level scheme for collecting and processing links as members of a link-base according to application specific requirements.

In hypermedia processing the context is an important concept. There are different contexts to recognise: The context a document appears or is to be presented in, the context of document fragments, given by its surrounding document data, and the context of a hyper reference. The latter decomposes into the source and destination context of a link, which more or less coincides with the context of the anchor fragments, and the context of the linking entity as discussed above.

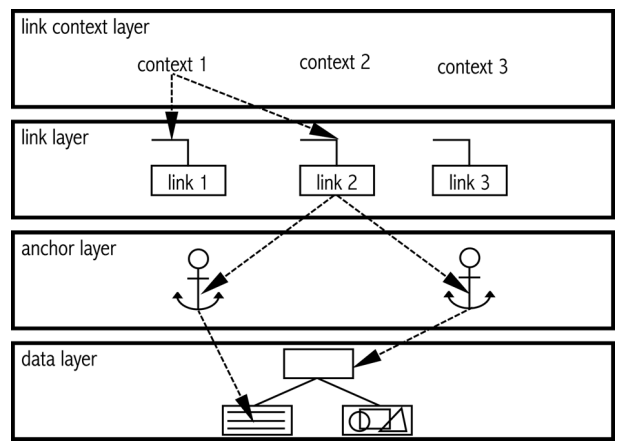

Figure 6: The MIRaCLE link layer model

In the present paper we are concerned with this semantically relevant link context. Since explicitly given by the author the context of a link need not to be determined at run time by analysing the current documents context. Link contexts are capable of articulating certain orthogonal information such as the author, the view or the proposed application of a hyper relation. To exploit these additional encodings a high-level semantic selection layer is needed to perform operations on link selections and collections based on the link context. Providing such mechanisms will enable users to steer hyperlink appearance by semantic criteria and thus interact more precise and purposeful with a hypermedia application. There are many imaginable operations like extracting links depending on their semantic role, attributes or on the relationship with their anchors or adapt them to users within personalised hypermedia applications.

The concept of a link context layer introduces a new abstraction on link collections. Within this layer we settle descriptions about a selection scheme for links following predefined semantic rules, operating on an abstract data model provided by the link layer. Link contexts neither create new links, nor new anchors. They are only responsible for the extraction of existing links stored in a link base. Those links are characterized by their descriptive properties as shown in the previous section, which combine in a selection scheme to represent a certain semantic context.

Link contexts are the upper tier in a four layered model consisting of a data, an anchor, a linking and the link context layer (s. figure 6). MIRaCLE (MIR adaptive context linking environment) is both, a formal model and a practical implementation based on the MIR system [3]. Each tier encapsulates certain data entities and access logics. All communication is only done between two neighbouring layers. The data layer stores raw data like text or images. Anchors are part of the anchor layer, addressing and marking chunks of data from the subordinate layer. Links connecting two or more anchors are provided by the link layer. Both, links and anchors are stored as discrete entities in a link base, offering addressability and meta data as part of an interface for the link context layer.

As shown before all semantically relevant notions from the link or anchoring layer are expressible in a formal RDF model. The link context itself is working on the model of the link layer representation and enables the user to select groups of links expressing a semantic relation. Retrieving links means picking sub-graphs from the model. The extraction of sub-graphs could be done by an appropriate query language, like RDQL [11]. Depending on the query language features one could take advantage of extended functionality like inference.

The result of such a query are statements which have the chosen links as subject and at least one predicate object pair formed by the involved anchors and the relationship existing between. Identifying the subject of the return statements as being a link gives all necessary information for further processing. An application could extract the participating anchors, verify them for being a start resource regarding the current document and visualize them, if wanted.

Thinking in our example on a vet student reading the text about hamster diseases let us imagine a context, which selects links providing some background information on the current topic like for example "Hamsters having hay fever". One possible link context definition is given by figure 7.

```
<?xml version="1.0"?>
<rdf:RDF xmlns:rdf="http://www.w3.org/1999/02/22-rdf-syntax-ns#"
xmlns:mir="http://www.rz.fhtw-berlin.de/MIR"
xmlns:dc="http://purl.org/dc/elements/1.1/">
<rdf:Description rdf:about="link-context1">
<dc:Creator>Mr. X</dc:Creator>
<dc:Title xmL:lang="en">Background Information</dc:Title>
<dc:Description xml:lang="en">Some continuative information
on.</dc:Description>
<mir:link-context>
<![CDATA[
SELECT * WHERE (?link, <rdf:predicate>, <mir:BackgroundInfo>) USING
rdf FOR <http://www.w3.org/1999/02/22-rdf-syntax-ns#>,
mir FOR <http://www.rz.fhtw-berlin.de/MIR#>
]]>
</mir:link-context>
</rdf:Description>
</rdf:RDF>
```

Figure 7: An example of a link context definition

The query will return all matching nodes in the graph which are the subjects of the associated RDF statements. The subjects will contain the name of the appropriate link and a statement about the connected anchors.

In the terms of our example it will return the link which expresses: "Link1 denotes that resource 'Hay fever handbook' presents BackgroundInfo to resource 'Hamsters having hay fever'". This higher order statement contains a simple statement embedding the target anchor being the subject, the predicate expressing the relation and the source anchor as the object. There are all necessary information for rendering the link into the document.

### 3.3 Implementing MIRaCLE on MIR

The MIR adaptive Context Linking Environment (MIRaCLE) was designed to meet the model sketched above. MIRaCLE is an adaptive scheme for dynamic link decoration and generation, especially suited for internet-based teaching. The system aims to enable teachers and students to define linking behaviour in semantic terms. MIRaCLE has been implemented within the MIR platform consisting of the following components:

o **Generic Anchors** are built from interfaces at all fine-grained information components. These anchor components are addressable without authoring effort.

o **Media-specific Anchors** are built upon the active reference features of the MIR system. Depending on the media type fragments of the data can be addressed with the help of mime-specific selector functions. For textual XML data the selector function is an XPointer/XPath implementation.

o **Links** are information entities, which provide references (to anchors) and metadata as needed to feed XLink expressions. Note that the MIR system automatically provides metadata exceeding the meager XLink standard (author, date, application context, …) and could provide an additional multitude without effort.

o **Link-base** resides within the MIR data store, which can be organised like a file system or grouped in application specific path spaces. The link-base offers an appropriate retrieval logic. Note that the MIR system automatically allows for structural traversal in downward and upward directions.

o **Link Context Layer** holds and processes the semantic selection instructions (RDQL queries) stored as link contexts within the MIR database. It uses the RDF model generated by the link layer as input and returns all RDF statements matching the given RDQL query. As the result of this "semantic filtering" the appropriate subset of links is forwarded to the page builder and page renderer.

All entities, anchors, links and the activated link contexts are processed on the fly as content gets observed through a standard Web browser.

## 4  Conclusions and Outlook

In this paper we discussed prospects and pitfalls of new educational content processing based on eLearning Objects, as derived from the IEEE standard LOM. Concepts and a practical solution for an open hypermedia ELO system were introduced. Starting from the available meta data set a translation into the semantic web framework was drawn. From these concepts a model for the semantic representation of hyperlinks was derived. With the introduction of semantic link contexts we arrived at a high-level scheme for generating and applying rhetorically consistent document coherence based on hyperlinks.

Much work, however, has to be done in this ongoing project. Our next steps will cover the application and the authoring of intelligent quizzing, which technically are easy to implement on MIR by means of an XSP logic library.

### Acknowledgements:

This work has been supported in part by the European Community within the Eumidis Pilot Project ODISEAME.

## Author(s):

Michael Engelhardt
FHTW Berlin, Hochschulrechenzentrum
Treskowallee 8, D-10318 Berlin, Germany
engelh@fhtw-berlin.de

Andreas Kárpáti
FHTW Berlin, Hochschulrechenzentrum
Treskowallee 8, D-10318 Berlin, Germany
karpati@fhtw-berlin.de

Torsten Rack
FHTW Berlin, Hochschulrechenzentrum
Treskowallee 8, D-10318 Berlin, Germany
rack@fhtw-berlin.de

Ivette Schmidt
FHTW Berlin, Hochschulrechenzentrum
Treskowallee 8, D-10318 Berlin, Germany
schmidti@fhtw-berlin.de

Thomas Schmidt, Dr.
FHTW Berlin, Hochschulrechenzentrum
Treskowallee 8, D-10318 Berlin, Germany
schmidt@fhtw-berlin.de